\documentclass[
]{ceurart}

\usepackage{listings}
\usepackage{colortbl}
\usepackage{amsmath}
\begin{document}

\copyrightyear{2024}
\copyrightclause{Copyright for this paper by its authors. Use permitted under Creative Commons License Attribution 4.0 International (CC BY 4.0)}
%\copyrightclause{Copyright for this paper by its authors.
 % Use permitted under Creative Commons License Attribution 4.0
%  International (CC BY 4.0).}

\conference{Forum for Information Retrieval Evaluation, December 12-15,2024,India}

\title{Enhancing Code Annotation Reliability: Generative AI’s Role in Comment Quality Assessment Models}

%%\tnotetext[1]{Forum for Information Retrieval Evaluation, December 15-18, 2023, India}

\author[1]{Seetharam Killivalavan}[%
orcid=0009-0002-2655-117X,
email=seetharam2210463@ssn.edu.in
]
\cormark[1]
\fnmark[1]
\address[1]{Sri Sivasubramaniya Nadar College of Engineering,
  Chennai, Tamil Nadu- 603110}
  
\author[2]{Durairaj Thenmozhi}[%
orcid=0000-0003-0681-6628,
email=theni$_d$@ssn.edu.in
]
\cormark[1]
\fnmark[1]

\cortext[1]{Corresponding author.}

\begin{abstract}
  This paper explores a novel method for enhancing binary classification models that assess code comment quality, leveraging Generative Artificial Intelligence to elevate model performance. By integrating 1,437 newly generated code-comment pairs, labeled as "Useful" or "Not Useful" and sourced from various GitHub repositories, into an existing C-language dataset of 9,048 pairs, we demonstrate substantial model improvements. Using an advanced Large Language Model, our approach yields a 5.78\% precision increase in the Support Vector Machine (SVM) model, improving from 0.79 to 0.8478, and a 2.17\% recall boost in the Artificial Neural Network (ANN) model, rising from 0.731 to 0.7527. These results underscore Generative AI's value in advancing code comment classification models, offering significant potential for enhanced accuracy in software development and quality control. This study provides a promising outlook on the integration of generative techniques for refining machine learning models in practical software engineering settings.

\end{abstract}


\maketitle

\section{Introduction}
Code comments are essential in software development, enhancing understanding, supporting team collaboration, and facilitating long-term code maintenance, as discussed by De et al. (2005) \cite{de2005study}. However, manually evaluating comments poses challenges due to its time-intensive and subjective nature, as noted by Haouari et al. (2011) \cite{haouari2011good}. To address these limitations, this study explores the use of Generative AI to automate comment quality assessment, as proposed by Ebert et al. (2023) \cite{ebert2023generative}, presenting a significant advancement for optimizing code review processes and expediting the Software Development Life Cycle (SDLC).

Incorporating comments effectively within the SDLC can benefit developers by accelerating troubleshooting, providing essential documentation, and establishing a robust groundwork for future development phases, as suggested by Majumdar (2020) \cite{majumdar2020comment}. This paper details our methods, experimental design, and the transformative potential of this AI-based approach for the software engineering field, as previously highlighted by Roehm et al. (2012) \cite{roehm2012professional}. Following this introduction, we review existing studies on comment classification and explain our process for generating a new dataset using Large Language Models (LLMs).

\subsection{	CODE COMMENT CLASSIFICATION: CURRENT LANDSCAPE AND CHALLENGES}

Code comments are used to clarify logic, design decisions, and develop challenges \cite{rani2021identify}. However, manual evaluation remains inconsistent, time-consuming, and subjective \cite{majumdar2020comment}. Automated classification, labeling comments as "Useful" or "Not Useful," offers a more efficient approach to streamline code review \cite{majumdar2022automated}. This study examines how Generative AI can enhance these classification models \cite{ebert2023generative}, potentially transforming comment quality assessment. By prioritizing essential comments, resource management can improve. This introduction sets up a discussion on how Large Language Models (LLMs) are advancing code comment classification and software development practices \cite{de2005study}.
\subsection{IMPACT OF LLM ON THE QUALITY OF COMMENTS}

Leveraging Large Language Models (LLMs) represents a major advancement in evaluating the quality of code comments \cite{ebert2023generative}. These models move beyond syntactic comprehension, capturing the deeper semantics of the code and generating insightful comments that streamline assessment processes. By doing so, they significantly enhance the relevance and clarity of comments across the Software Development Life Cycle (SDLC). Beyond mere classification, LLMs redefine developer interaction with code, fostering clearer communication and strengthening collaboration. This transformative impact underscores the essential role LLMs are set to play in the future of code comment quality evaluation.

\vspace{\baselineskip}
{
\raggedright
The application of Generative AI within the IRSE@FIRE-2024 task \cite{majumdar2023generative} is set to transform code quality evaluation, streamlining the Software Development Life Cycle (SDLC) and promoting more effective resource distribution and collaborative development efforts among teams.
\vspace{\baselineskip}

The subsequent sections are organized as follows: Section \ref{sec2} provides an overview of comment classification and the foundations of Generative AI. Section \ref{sec3} describes the task setup and dataset used. Our methodology is detailed in Section \ref{sec4}. In Section \ref{sec5}, we present the results, while Section \ref{sec6} offers a comparative analysis of our models and embeddings against established approaches in code comment quality assessment, underscoring their unique contributions. Lastly, Section \ref{sec7} concludes with a summary of our findings and discusses possible avenues for future research.
}

\section{Related Work \label{sec2}}
Automated program understanding is a recognized research area among professionals in the software domain.
Various tools have been developed to facilitate the extraction of knowledge from software metadata, encompassing components such as runtime traces and structural attributes of code~\cite{de2005study, majumdar2019smartkt, chatterjee2015debugging, majumdar2016d, majumdar2021mathematical, majumdar2021dcube_, siegmund2017measuring, chatterjee2023parallelc}.
Researchers have developed various methods to mine and evaluate code comments, focusing on analyzing comment quality through code-comment pair comparisons. In assessing code comment quality, authors~\cite{tan2007hotcomments,wang2023codet5+,steidl2013quality,majumdar2022can,majumdar2022overview,freitas2012comment,majumdar2023smart,chakraborty2023bringing} employ techniques such as word similarity measures (e.g., Levenshtein distance) and comment length analysis to filter out trivial and non-informative comments. Rahman et al.~\citep{rahman2017predicting} detect useful and non-useful code review comments (logged-in review portals) based on attributes identified from a survey conducted with developers of Microsoft~\citep{bosu2015characteristics}.

New programmers often rely on existing comments to comprehend code flow. However, not all comments contribute effectively to program comprehension, necessitating a relevancy assessment of source code comments prior to their use. Numerous researchers have focused on the automatic classification of source code comments in terms of quality evaluation. For instance, Omal et al.~\cite{oman1992metrics} noted that factors influencing software maintainability can be organized into hierarchical structures. The authors defined measurable attributes in the form of metrics for each factor, enabling the assessment of software characteristics, which can then be consolidated into a single index of software maintainability. Fluri et al.\cite{fluri2007code} examined whether the source code and associated comments are changed together along the multiple versions. They investigated three open source systems, such as {\em ArgoUML, Azureus}, and {\em JDT Core}, and found that 97\% of the comment changes are done in the same revision as the associated source code changes.  Yu Hai et al.\cite{yu2016source} classified source code comments into four classes - unqualified, qualified, good, and excellent. The aggregation of basic classification algorithms further improved the classification result. Another work published in \cite{majumdar2022automated} in which author proposed an automatic classification mechanism "CommentProbe" for quality evaluation of code comments of C codebases. We see that people worked on source code comments with different aspects\cite{majumdar2022automated,majumdar2020comment,majumdar2022overview,majumdar2022can,majumdar2023smart,chakraborty2023bringing}, but still, automatic quality evaluation of source code comments is an important area and demands more research.

With the advent of large language models~\cite{brown2020language}, it is important to compare the quality assessment of code comments by the standard models like GPT 3.5 or llama with the human interpretation. The IRSE track at FIRE 2024~\cite{paul2024generative,paul2024overview} builds upon the methodologies proposed in~\cite{majumdar2022automated,paul2023efficiency,majumdar2023generative,majumdar2022can} to investigate various vector space models~\cite{majumdar2022effective} and features for binary classification and evaluation of comments in relation to code comprehension. This track also assesses the performance of the predictive model by incorporating GPT-generated labels for the quality of code and comment snippets extracted from open-source software.

\section{Task and Dataset Description \label{sec3}}
This section outlines the IRSE@FIRE-2024 task \cite{majumdar2023generative}, focused on improving a binary code comment quality classification model. The task involves integrating newly generated code-comment pairs for enhanced accuracy. It comprises an initial dataset of 9048 labeled code-comment pairs in C, out of which 5378 were classified as "Useful" and 3670 were classified as "Not Useful", along with additional pairs generated using a Large Language Model (LLM), each labeled.

The desired output includes two versions of the classification model: one with the added generated pairs and labels, and another without. The starting dataset encompasses 9048 comments from GitHub, each with the comment text, surrounding code, and a corresponding usefulness label (Table \hyperlink{topoftab1}{1}).

\begin{table}[ht]
\hypertarget{topoftab1}{}
\caption{Sample Data Instance}
\centering
\label{tab1}
\begin{tabular}{|p{5cm}|p{5cm}|p{2.5cm}|}
\hline
{\cellcolor[rgb]{0.753,0.753,0.753}}\textbf{Comment}& {\cellcolor[rgb]{0.753,0.753,0.753}}\textbf{Code}& {\cellcolor[rgb]{0.753,0.753,0.753}}\textbf{Label}\\
\hline
/* Swap two values */& void swapValues(int *x, int *y) \{

 int temp; 

temp = *x; 
 *x = *y; 
  
  *y = temp;\}& Useful \\
\hline
/* Compute the sixth power */ & {int result = computeSixthPower(value);}& Not Useful \\
\hline
/*Simple variable declaration */& 
int x = 10;
& Not Useful \\
\hline
\end{tabular}

\end{table}

To establish the ground truth, 14 annotators assessed each comment independently, resulting in substantial agreement (Cohen’s kappa value of 0.734). The annotation process involved the assessment of a comprehensive set of  16,000 comments.

Participants are also tasked with generating an additional dataset of labeled code-comment pairs from GitHub using an LLM. This dataset is to be submitted alongside the task.

In summary, the objective is to refine the code comment quality classification model by integrating newly generated pairs, ultimately enhancing accuracy and effectiveness.

{
\raggedright
For further details, please refer to the task description provided at IRSE@FIRE-2024 \footnote{https://sites.google.com/view/irse2024/home}.
}

\section{Methodology \label{sec4}}
Our approach encompasses the combination robust methodologies, including Support Vector Machine (SVM) models for classification and Artificial Neural Networks (ANN) with diverse activation functions for capturing complex data relationships \cite{igual2017introduction}. Additionally, we leverage Large Language Models (LLMs) via the OpenAI API and utilize GitHub repositories to generate a diverse and substantial dataset of code-comment pairs. The following subtopics detail our specific methodologies: implementing SVM models, exploring ANN models, and generating datasets using the OpenAI API and GitHub repositories. These methodologies collectively form the foundation of our innovative approach to code comment quality assessment. Within the framework of our methodology, Figure \hyperlink{topofimg1}{1} elegantly elucidates the architectural blueprint that underpins our approach.

\begin{figure}[ht]
\hypertarget{topofimg1}{}
\centering
\includegraphics[width=\linewidth]{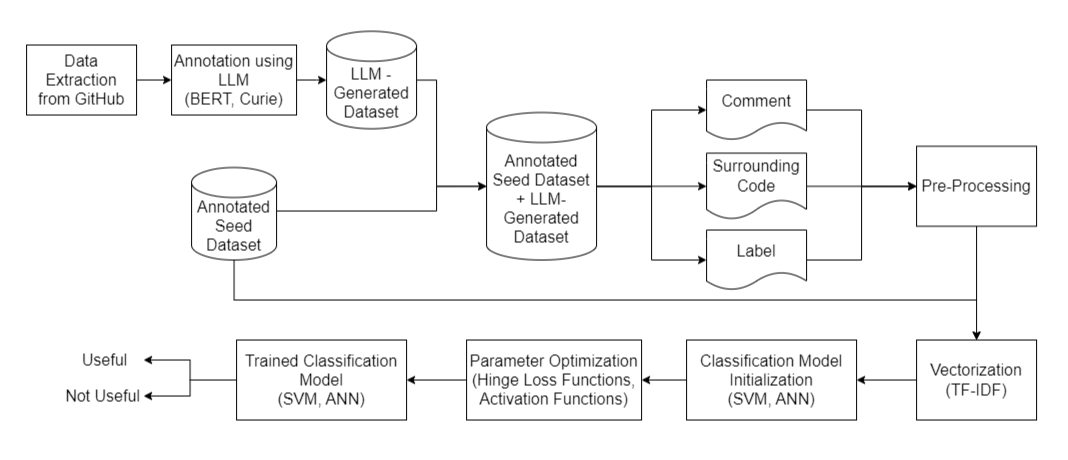}
\caption{Architecture Diagram}
\label{fig1}
\end{figure}

\subsection{Support Vector Machines}

A Linear Support Vector Machine (SVM) is a powerful classification technique that finds the optimal hyperplane for effective data separation, expressed as \(y = mx + b\), where \(y\) is the predicted class label, \(x\) is the input data, \(m\) is the slope and \(b\) is the y-intercept. It maximizes the margin, which is the distance between the hyperplane and the nearest data points. This margin (\textit{M}) can be calculated as:

\begin{equation}
M = \frac{2}{\|m\|} \tag{1}
\end{equation}

where ||\textit{m}|| is the length of the weight vector\textit{ m}.

{
\raggedright
SVM aims to minimize the square of the length of the weight vector (||\textit{m}||²) while ensuring that each data point $x_i$ is correctly classified:
}

\begin{equation}
y_i (m \cdot x_i + b) \geq 1 \tag{2}
\label{eqn2}
\end{equation}

{
\raggedright
Equation \ref{eqn2} states that the product must be greater than or equal to 1 for all data points, emphasizing the importance of well-defined class separation in SVM classification. This condition is central to SVM's goal of locating an optimal hyperplane, maximizing the margin, and guaranteeing accurate data point classification. Support vectors, those closest to the hyperplane, are pivotal in margin definition, thereby influencing SVM's overall performance.
}

\subsection{Artificial Neural Networks}
Artificial Neural Networks (ANNs) are adaptable machine learning models that draw inspiration from the architecture and operation of the human brain. They excel at discerning complex data relationships, making them highly effective for tasks like code comment quality classification.
The mathematical representation of a single neuron in an ANN is given by:

\begin{equation}
    Z = \textit{w}_1\textit{x}_1 + \textit{w}_2\textit{x}_2 + \ldots + \textit{w}_n\textit{x}_n + \textit{b} \tag{3}
\end{equation}

where $x_n$ are input features, $w_n$ are corresponding weights and $b$ is the bias term.

{
\raggedright
\vspace{\baselineskip}
The weighted sum (\textit{Z}) is then passed through an
activation function, which introduces non-linearity into the model. Different
activation functions yield different learning behaviours.

\vspace{\baselineskip}

Here are a few common activation functions and their
formulas:
\vspace{\baselineskip}
}

i) Logistic Function:
\[f(Z) = \frac{1}{1 + e^{-Z}} \tag{4}\]

ii) Rectified Linear Unit (ReLU):
\[f(Z) = \max(0, Z) \tag{5}\]

iii) Hyperbolic Tangent (tanh):
\[f(Z) = \frac{e^Z - e^{-Z}}{e^Z + e^{-Z}} \tag{6}\]
\subsection{Leveraging LLM for Generation of Dataset}
Our methodology encompasses a multi-step approach to dataset generation. Initially, we leveraged both the OpenAI API, powered by the Curie Model, and GitHub repositories to diversify our dataset. The API simulated real-world coding scenarios, producing authentic code-comment pairs and substantially augmenting our dataset. Complementing this, we extracted additional pairs from various open-source projects on GitHub, ensuring relevance and utility. This combined strategy significantly broadened the dataset's coverage while upholding high quality standards. Subsequently, the code-comment pairs underwent processing using OpenAI's Curie Model in conjunction with BERT for label generation, signifying comment usefulness. This involved presenting prompts with both code and comment, and employing the LLM to generate a label. Finally, the dataset was meticulously assembled, each entry comprising code, comment, and the corresponding generated label. This rigorous methodology serves as a robust foundation for our code comment quality classification model.

\section{Analysis of Results \label{sec5}}
Evaluating our code comment quality classification model is a crucial step in validating its effectiveness. We utilized a combination of Support Vector Machines (SVM) and Artificial Neural Networks (ANN) with various activation functions, including ReLU, identity, logistic, and tanh, to conduct a comprehensive analysis of the model’s performance. This multidimensional approach offered valuable insights into the model's adaptability, revealing its robustness across diverse scenarios. Additionally, integrating these methodologies resulted in a significant improvement in precision, underscoring the model’s ability to categorize code comments accurately based on practical value. These findings align with previous research that demonstrates the reliability of SVM and ANN models for comment quality assessment. The use of diverse activation functions further highlights the flexibility of our approach, reinforcing the model’s potential applicability in real-world software development.
\subsection{Classification Models}
The evaluation of our code comment quality classification models yielded insightful findings, showcasing the impact of integrating LLM-generated data into our seed dataset of 9048 entries. This initial dataset was thoughtfully partitioned into training, testing and validation sets, with the testing set comprising 1718 entries. 
With the Seed Data,  SVM exhibited commendable precision (0.79), while ANN with ReLU activation demonstrated remarkable effectiveness, resulting in a notable recall score (0.731). Models with tanh and logistic activation functions showed similar precision scores of 0.726 and 0.73.

Post integration of 1437 LLM-generated entries, which seamlessly enriched the Seed Data, SVM's precision notably increased by 5.78\%, elevating the preceding value to 0.8478, highlighting the value of incorporating generative AI. Using ReLU, ANN achieved a noteworthy 2.17\% rise in its recall, giving it a final recall of 0.7527, while tanh and logistic functions yielded marginal changes. Extensive experimentation with varied SVM models and ANN activation functions was performed, and the results depicts the effectiveness of our approach, emphasizing the importance of meticulous experimentation in fine-tuning models for code comment quality analysis.

Furthermore, for detailed numerical insights, please refer to Table \hyperlink{topoftab2}{2}, which provides a comparison of the model performance, offering the classification report of our top-performing models. It serves as a comprehensive reference for our findings and allows the comparison of test accuracies and F1 scores before and after integration.

\vspace{\baselineskip}

\begin{table}[ht]
\hypertarget{topoftab2}{}
\label{tab2}
\centering
\caption{Model Performance Comparison}
\begin{tabular}{|c|c|c|c|c|}
\hline
\rowcolor[rgb]{0.753,0.753,0.753}
\textbf{Model}& \multicolumn{2}{c|}{\textbf{Performance with Seed Data}} & \multicolumn{2}{c|}{\textbf{Performance with Integrated Data}} \\
\cline{2-5}
\rowcolor[rgb]{0.753,0.753,0.753} & \textbf{Test Accuracy} & \textbf{F1-Score} & \textbf{Test Accuracy} & \textbf{F1-Score} \\
\hline
Linear SVM& 0.811& 0.792& 0.813& 0.79\\
\hline
SVM (poly. kernel) & 0.798& 0.781& 0.814& 0.796\\
\hline
ANN (ReLU) & 0.75& 0.741& 0.757& 0.744\\
\hline
ANN (tanh) & 0.753& 0.729& 0.749& 0.743\\
\hline
ANN (logistic) & 0.749& 0.73& 0.745& 0.736\\
\hline
ANN (identity) & 0.748& 0.728& 0.743& 0.729 \\
\hline
\end{tabular}
\end{table}

\subsection{Analysis of Dataset Generated using LLM}
The integration of data generated by OpenAI's Large Language Model (LLM), in conjunction with the utilization of the Curie model, and the inclusion of diverse datasets from various GitHub repositories and open-source projects represents a significant stride in elevating our code comment quality classification model. By meticulously adding 1437 new entries to our original dataset, we substantially enriched the diversity of our training corpus. This augmentation in data diversity led to a marked improvement in the accuracy of our classification model, benefiting both Support Vector Machine (SVM) and Artificial Neural Network (ANN) models. The heightened sensitivity achieved through this amalgamation enhances the model's generalization and prediction capabilities, underscoring the value of incorporating external data sources. Furthermore, the integration of BERT embeddings and the Curie model empowered our model to adeptly capture the intricacies of code commentary, notably enhancing its ability to distinguish between "Useful" and "Not Useful" comments. This capability proves crucial in real-world scenarios, where precise comment assessment plays a pivotal role in influencing the effectiveness of software development and maintenance processes.

\section{Discussion \label{sec6}}
In this section, we conduct a thorough comparative analysis of our models and embeddings in relation to previous studies on code comment classification. Our deliberate emphasis on Support Vector Machine (SVM) and Artificial Neural Network (ANN) models, each with specific activation functions, allows for an in-depth exploration of their efficacy. This focused investigation provides nuanced insights into their performance in code comment quality assessment, contrasting with the broader set of classifiers utilized by Majumdar et al. (2022a) \cite{majumdar2022automated}.

Additionally, our research methodology diverges from the work of Majumdar et al. (2020) \cite{majumdar2020comment}, which primarily centers on the extraction of knowledge domains from code comments for addressing developer queries during maintenance. In contrast, our focus centers on the development and evaluation of code comment quality classification models. This includes the integration of LLM-generated data, resulting in significant enhancements in classification precision.

Concerning embeddings, Majumdar et al. (2022b) \cite{majumdar2022effective} emphasize contextualized word representations fine-tuned on software development texts. In our case, we utilized both BERT and custom embeddings specifically tailored for software development concepts. This approach provided high-dimensional semantic representations, catering to a wide array of natural language processing tasks. It's worth noting that for labeling, we harnessed the Curie model. This distinction underscores the versatility and broader applicability of our embeddings compared to the contextualized embeddings discussed by Majumdar et al (2022b)\cite{majumdar2022effective}.

Fundamentally, our proposition emphatically focuses on specific models and embeddings, providing unique insights into their effectiveness for assessing code comment quality. The emphasis on specific models and customized embeddings offers detailed insights into evaluating code comment quality, distinguishing it from the broader, contextually-focused techniques utilized in prior research \cite{majumdar2022effective}.

\section{Conclusion \label{sec7}}
Building on these foundational advancements, our study highlights the practicality and scalability of Generative AI for real-world applications. By generating and integrating new data into existing datasets, we demonstrated that Generative AI could enhance the performance of traditional models in code comment quality assessment. This approach not only elevated our models’ precision and recall but also underscored the potential of Generative AI to provide robust solutions for improving software documentation practices, making it an impactful tool for future development cycles.

\vspace{\baselineskip}
{

The integration of LLM-generated data notably amplified model performance, with precision for the SVM model increasing by 5.78\% and recall for the ANN model improving by 2.17\%. These enhancements raised the test accuracies to 81.1\% for SVM and 75\% for ANN, marking a clear advancement from their pre-augmentation baselines. These quantifiable gains underscore the effectiveness of data augmentation via Generative AI, illustrating how even modest dataset expansions can yield substantial improvements in model accuracy and reliability, particularly for complex classification tasks in software development.

\vspace{\baselineskip}

Looking ahead, the impact of this work can extend well beyond code comment classification. The methodologies introduced here establish a versatile framework that can be adapted for a wide range of tasks in software development and quality assurance. By leveraging generative AI, specifically through Large Language Models (LLMs), we highlight a powerful approach that could redefine code analysis and documentation practices. As the software industry evolves, this research stands as evidence of the substantial value in adopting advanced technologies, reinforcing the importance of innovative solutions in enhancing efficiency and precision in practical engineering applications.
}

%%
%% Define the bibliography file to be used
\bibliography{sample-ceur}


\end{document}